\documentclass[iop]{emulateapj}
\usepackage{color}
\usepackage{natbib}
\slugcomment{}

\shorttitle{Chemistry of distinct RGBs in NGC~6752}
\shortauthors{Carretta et al.}

\begin{document}
\title{Chemical tagging of three distinct populations of red giants in the
globular cluster NGC~6752\altaffilmark{1}}

\author{E. Carretta\altaffilmark{2},
A. Bragaglia\altaffilmark{2},
R.G. Gratton\altaffilmark{3},
S. Lucatello\altaffilmark{3},
V. D'Orazi\altaffilmark{4}
}

\altaffiltext{1}{Based on data collected at the ESO telescopes under 
programme 085.D-0205}
\altaffiltext{2}{INAF, Osservatorio Astronomico di Bologna, via Ranzani 1,
       40127,  Bologna,  Italy. eugenio.carretta@oabo.inaf.it,
       angela.bragaglia@oabo.inaf.it}
\altaffiltext{3}{INAF, Osservatorio Astronomico di Padova, vicolo
       dell'Osservatorio 5, 35122 Padova,  Italy. raffaele.gratton@oapd.inaf.it
       sara.lucatello@oapd.inaf.it}
\altaffiltext{4}{Dept. of Physics and Astronomy, Macquarie University, Sydney, 
NSW, 2109 Australia. valentina.dorazi@mq.edu.au}

\begin{abstract}

We present aluminium, magnesium, and silicon abundances in the metal-poor globular
cluster NGC~6752 for a sample of more than 130 red giants with homogeneous
oxygen and sodium abundances. 
We find that  [Al/Fe] shows a spread of about 1.4 dex among giants in
NGC~6752 and is anticorrelated with [Mg/Fe] and [O/Fe]  and correlated 
with [Na/Fe] and [Si/Fe]. These relations are not continuous in nature,
but the distribution of stars is clearly clustered around three distinct Al
values, low, intermediate, and high. These three groups nicely
correspond to the three distinct sequences previously detected using Str\"omgren
photometry along the red giant branch. These two independent findings strongly
indicate the existence of three distinct stellar populations in NGC~6752.
Comparing the abundances of O and Mg, we find that the population with
intermediate chemical abundances cannot originate from material with the same
composition of the most O- and Mg-poor population, diluted by material with that
of the most O- and Mg-rich one. This calls for different polluters.

\end{abstract}


\section{Introduction}

Efficient multi-object high-resolution spectrographs allowed in
recent years a major step forward in understanding multiple
stellar populations in globular clusters (GCs). The large star-to-star abundance
variations of light elements (C, N, O, F, Mg, Al, Si) were found to reflect the
composition of at least two distinct stellar generations (e.g., \citealt{gra01}
and the reviews by \citealt{gra04}, \citealt{gra12}). The huge gain in
statistics not only provides a striking improvement for individual clusters, but
better data available for large samples in a large number of GCs allow to
distinguish cluster-to-cluster differences in the extension and shape of the
related anticorrelations. 

A good fraction of the observed stars comes from our ongoing FLAMES survey 
(\citealt{car06,car09a,car09b} for a description), devoted to study the Na-O
anticorrelation and its relation to global parameters in a large
sample of GCs. Quantifying its extension, we were for instance able
to show its dependence on the total absolute magnitude and metallicity of the
clusters \citep{car10} as well as its correlation with helium abundance
\citep{gra10}. 

Better insight on the (still debated) nature of candidate polluters can be
obtained by coupling the NeNa and the MgAl cycles. A good test-bed comes again
from our FLAMES survey, and consists in more than 200 stars with FLAMES-UVES
spectra, from which we derived simultaneously O, Na, Mg, and Al abundances
\citep{car09b}. The results show how in some clusters (usually massive and/or
metal-poor, such as NGC~2808, NGC~6752) large amount of Al -and  likely He- are
produced, with efficient destruction of O and Mg, whereas in others (e.g. M~4)
no fresh He and Al is present.  The major drawback is the limited sample
available for each cluster (up to 14 stars observed with UVES Red Arm). To
overcome this problem, we observed some interesting GCs with
FLAMES-GIRAFFE to add Al abundances to the  large set of O, Na, and Mg
abundances already in our hand.

These observations may furthermore shed some light on the open issue of the
continuous vs discrete nature of multiple stellar generations in GCs. At
present, only in the case of the discrete main sequences in NGC~2808 the
chemical signature of the Mg-Al anticorrelation was found, directly studying a
He-rich and a He-poor dwarf with X-shooter \citep{bra10} at intermediate
resolution.  In the other case of clearly separated main sequences, $\omega$
Cen, it was only possible to infer a very He-enriched distinct sequence using
stacked spectra of similar resolution \citep{pio07}.
High resolution spectra of a large sample of stars (e.g.\citealt{mar08} in M~4,
with UVES) offer a better window on possible clumped distributions along the
Na-O anticorrelation, but only at a large cost in observing time (about 30 hours
with UVES versus about 30 minutes with GIRAFFE of \citealt{car09a}). 
Hence, we exploited the efficient GIRAFFE HR21 grating to get a deeper
insight on these two aspects: what were the first generation polluters (either
intermediate-mass AGB stars, \citealt{ven01}, or fast rotating massive stars,
FRMS, \citealt{decr07}), and whether the bursts of star formation for
the different stellar generations in GCs represented a continuum or were
discrete in nature.
In this Letter we present the first results for NGC~6752 ([Fe/H]=-1.55,
$M_V=-7.73$, \citealt{car10}), showing that our
chemical tagging strongly confirm that three distinct populations are clearly
traceable along the red giant branch (RGB) in this metal-poor cluster for which
\cite{mil10} only found some evidence of an intrinsic broadening of the
main-sequence.
We briefly discuss the implications that the compositions of the three
 populations  have on our understanding of  the nature of the polluters.

\section{Data and analysis}

The abundances of aluminium were derived  from
GIRAFFE spectra (obtained with FLAMES, \citealt{pas02}, mounted at the ESO
VLT-UT2 telescope) with the high resolution grating HR21. This set-up is centred
at 8757~\AA\ and the spectral resolution is R=17,300 at the center of spectra.
The pointing was made adopting the same fibre positioning used by \cite{car07}
to observe NGC~6752 with HR11, which includes the Na doublet at 5682-5688~\AA:
this choice maximizes the number of stars with both Na and Al abundances
available\footnote{O abundances, obtained using the HR13 grating and a
slightly different fibre configuration, are not available for all the program 
stars}. The observations were made on July 2010, with an exposure time of
2700 seconds, and reduced by the dedicated GIRAFFE pipeline. The resulting 1-D,
wavelength-calibrated spectra were sky subtracted, and shifted to zero radial
velocity. The S/N values span a range from about 50 up to more than 500, with
a median value of 192.

Equivalent widths ($EW$s) of the Al lines were measured with the ROSA package
\citep{gra88} and abundances were derived using the atmospheric parameters
already determined for each star in \cite{car07}. A check with spectrum
synthesis confirmed the reliability of our measurements. We use here LTE
abundances. \cite{andrievski} computed N-LTE
corrections for Al for very metal-poor stars, considering also the doublet used
here. From their Fig.~2 the maximun differential effect for stars with
temperature and gravity ranges similar to ours is less than 0.2 dex for
[Fe/H]=-2, the highest metallicity they consider, and arguably less at the
cluster metallicity; futhermore, there is no trend of Al abundances with
temperature or gravity in our data. 
The Mg abundances rest on two to three high
excitation lines measured in the spectral ranges of HR11 and HR13, while Si was
obtained from several transitions in the spectral range 5645-6145~\AA ~(details
and atomic parameters can be found in \citealt{car09b} and \citealt{gra03},
respectively). The uppermost left panel in Fig.~\ref{f:fig1} shows the comparison of
our [Al/Fe] ratios derived from the 8772-8773~\AA\ doublet with those obtained
by \cite{yon05} using the weaker doublet of Al~{\sc i} at
6696-98~\AA\ in UVES spectra with spectral resolution ranging from R=60,000 to
110,000 and with S/N ratios from 100 to 250 per pixel. The agreement is
excellent: our abundances are on average only 0.01 dex lower than those from
this other study, with an $r.m.s.$ scatter of 0.14 dex (13 stars). Abundances
of Mg, Al, and Si for individual stars are listed in Table 1. Typical
star-to-star errors in abundance ratios, due to errors on the adopted
atmospheric parameters and $EW$ measurements, were estimated as in \cite{car07};
they are 0.07, 0.10 and 0.05 dex for [Al/Fe], [Mg/Fe] and [Si/Fe], respectively.
 
\section{Aluminum, magnesium, and silicon}

\begin{figure} 
\centering
\includegraphics[bb=26 144 424 719, scale=0.80]{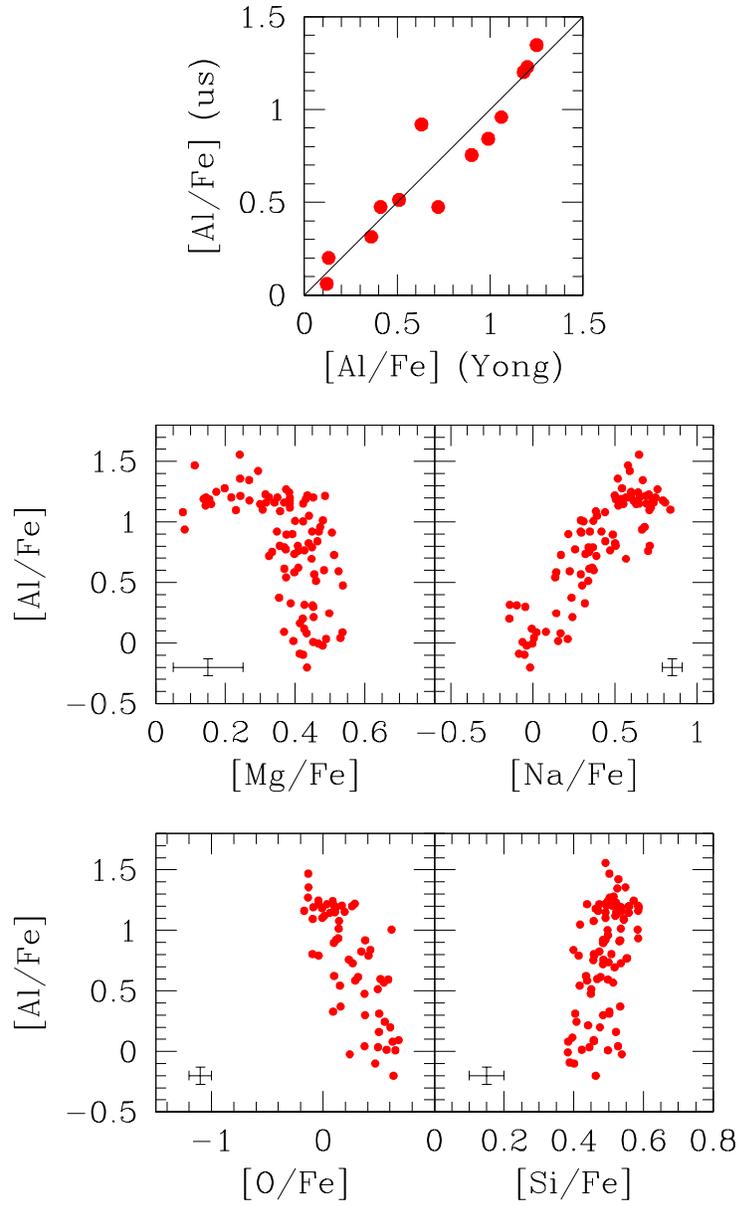}
\caption{Upper panel, left: comparison of our [Al/Fe] ratios with
those from \cite{yon05} for 13 stars in common. The line of identity is
indicated. Upper panel, right: the Mg-O correlation from our data (present work
and \citealt{car07}). Middle panels: the Al-Mg anticorrelation (left) and the
Na-Al correlation (right). Lower panels: the same for the Al-O anticorrelation
(left) and the Al-Si correlation (right).}
\label{f:fig1}
\end{figure}

In the middle and lower panels of Fig.~\ref{f:fig1} we show the relations 
between the abundances of Al and of other proton-capture elements Na, O,
Mg, and Si in NGC~6752. We observe large star-to-star variations in the Al
content, with a range of about 1.4 dex. The stars with the minimum (i.e.,
primordial) Al abundance also show the whole pattern typical of core collapse SN
nucleosynthesis only: [Na/Fe]$\sim -0.1$ dex, [Mg/Fe] and [Si/Fe]$\sim +0.45$
dex, [O/Fe]$\sim +0.5$ dex. The stars with most extreme abundances reach values
of [Al/Fe] as large as 1.4-1.5 dex. This observation by itself is a  severe
challenge to models invoking AGB stars as first generation polluters, since the
only models able to produce such extreme abundances of Al \citep{ven09} are
those of metal-poor ([Fe/H]$\sim -2.3$), low-mass (3.0-3.5 $M_\odot$) AGBs.
However, these models do not show a depletion, but rather a simultaneous
$overproduction$ of O. Moreover, stars in this mass range should experience a
prolonged phase of 3rd-dredge up, likely resulting in enhanced C and
$s-$elements, which is not observed in NGC~6752 \citep{jam04}. On the other
hand, models of  5 $M_\odot$ stars at the metallicity of NGC~6752 do yield
abundances [Al/Fe]$\sim 1.2$ dex, not very far away from the observed maximum,
but they fail to reproduce the whole range of enhancements observed in Na (see
Table 2 in \citealt{ven09}). Finally, even the nucleosynthesis of
rapidly-rotating massive stars cannot  reproduce such high Al abundances (see
Decressin et al. 2007).

As expected, Al is well correlated with elements that are enhanced by the
action of the Ne-Na (such as Na) and Mg-Al cycles and is anticorrelated with
elements that are depleted in H-burning at high temperature (such as O and Mg). 
Furthermore, with the present work we confirm that in NGC~6752 the reaction network
within the polluters of the first generation exceeded the temperature of $\sim
65\times 10^6$ K , the threshold  required to generate the leakage from the MgAl
cycle on $^{28}$Si, first discovered by \cite{yon05} in NGC~6752. The resulting
Al-Si correlation was then confirmed by \cite{car09b} using a limited sample of
stars with UVES spectra, and is visible also in the lower-right panel in
Fig.~\ref{f:fig1}, based on a much more extended sample. This correlation is
typically observed in metal-poor and/or massive clusters, like NGC~2808 and 
NGC~7078 (M~15; see \citealt{car09b}), and is a clear cut signature for the
action of a class of first generation polluters so massive to reach such high
temperature in their interiors. 

\section{Three distinct sequences on the RGB in NGC~6752}

\begin{figure} 
\includegraphics[scale=0.80]{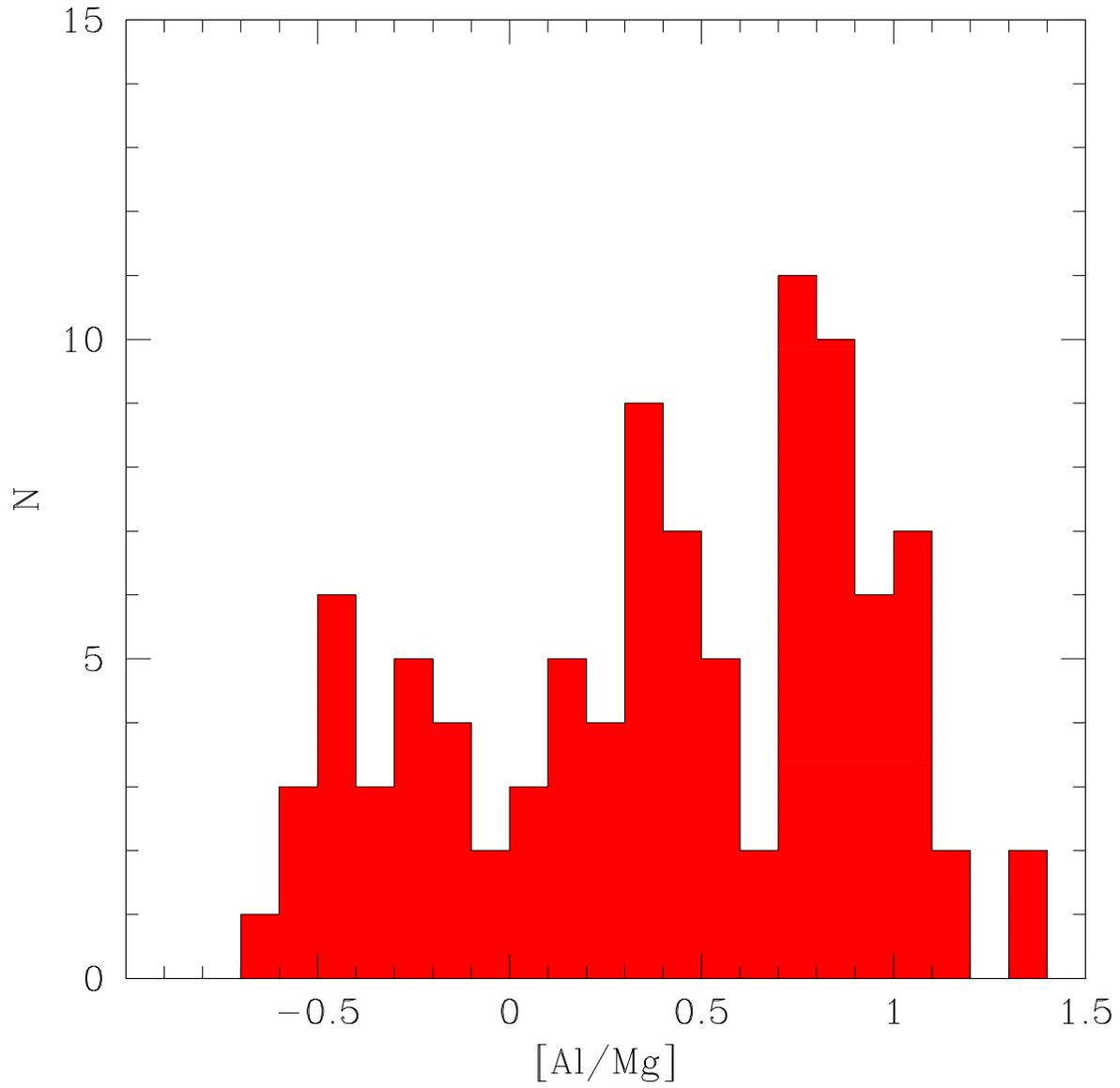}
\caption{Distribution of the [Al/Mg] ratios in red giants in NGC~6752}
\label{f:fig2}
\end{figure}

The most striking result visible in the present large dataset is however the
clumpy appearance of the distributions displayed in Fig.~\ref{f:fig1}. Three
groups of stars appear to be clustered around [Al/Fe] values of about 0.1, 0.8,
and 1.2 dex, respectively. This division seems reminiscent of that into the
primordial (first generation, P) and intermediate (I) and extreme (E) components
of second generation defined in \cite{car09a} using Na and O abundances. To
better distinguish the three subgroups, we plot in Fig.~\ref{f:fig2} the
distribution of the [Al/Mg] values for red giants in NGC~6752. This ratio
actually maximizes the signal along the anticorrelation and allows us to nicely
trace three distinct groups with gaps at [Al/Mg]$\sim 0.0$ dex and [Al/Mg]$\sim
0.65$ dex. In turn, these gaps correspond quite well to the breaks clearly
visible in the Al-Mg anticorrelation at [Al/Fe]$=0.4$ and [Al/Fe]$\sim 1.0$ dex,
both at [Mg/Fe]$\sim 0.45$ dex.

We identify the first break as a well defined border at the end
of the star formation of the first stellar generation in the cluster. The
novelty is that the gap between the groups of second generation stars with
intermediate and extreme composition is now clearer than what we see
using the Na-O anticorrelation in the same stars. In the large dataset
by \cite{car07} there is not a clear distinction between these groups; on the
other hand, in the combined sample shown by \cite{yon03} (their Fig.4) three
groups stand out in the Al-Na plane.

How solid is our finding? We can check this by using a fully independent
approach. In \cite{car11} we used Str\"omgren multiband photometry, combined
with information from spectroscopy, to study the possible separation and the
properties of multiple populations in GCs. We tried to find the best combination
of filters to enhance the differences (due to N and C molecular bands) and
defined a  new index $\delta_4 = (u-v)-(b-y)$. This index seems to be better
suited than others (such as $c_y$ by \citealt{yon08}) at separating
first and second generation GC stars in the whole metallicity range sampled by
GCs (see \citealt{car11} for details). 

\begin{figure} 
\includegraphics[bb=18 362 582 705, scale=0.80]{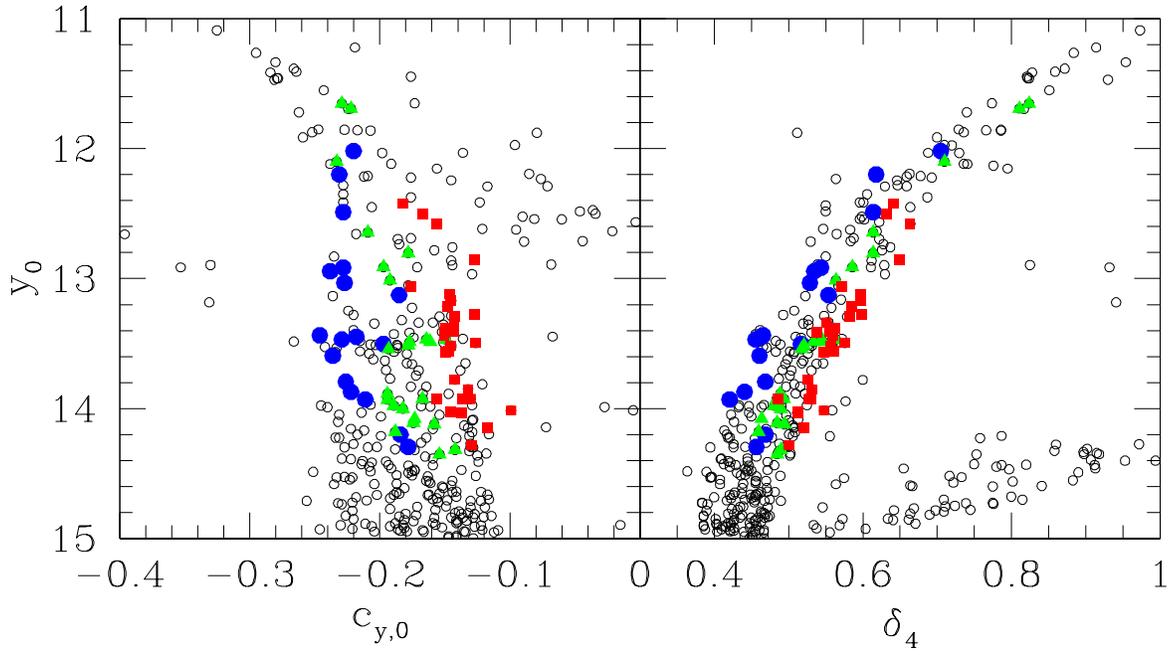}
\caption{Left panel: Str\"omgren (dereddened) CMD in $y,c_y,$ for NGC~6752
(empty grey circles). Superimposed in blue circles, green triangles and red 
squares are the
stars with measured Al abundances: [Al/Fe]$<0.4$ dex, $0.4<$[Al/Fe]$<1.05$ dex
and [Al/Fe]$>1.05$ dex, respectively. Right panel: the same for the CMD in
$y,\delta_4$, where $\delta_4$ is a new index defined as $\delta_4=(u-v)-(b-y)$
(see \citealt{car11}).}
\label{f:fig3}
\end{figure}

Anyway, both indexes are able to show distinct structures on the RGB of
NGC~6752, as demonstrated by Fig.~\ref{f:fig3}: NGC~6752 displays a tri-modal
distribution, well evident both in the $y_0,c_{y,0}$ and in $y_0,\delta_4$
color-magnitude diagrams (CMD). The conclusion from this sample of hundreds of
stars is that in NGC~6752 there have been three episodes of star formation with
well mixed gas in each individual phase.

Taking advantage of our newly derived abundances of Al and Mg for about 130
giants, we are now in the position to check this inference. We have about 110
stars in common with the sample of giants of NGC~6752 with accurate Str\"omgren
photometry. In Fig.~\ref{f:fig3} we indicate   stars from the present analysis,
dividing them in three different groups, according to their Al abundances: stars
with [Al/Fe]$<0.4$ dex (blue filled symbols), stars with $0.4<$[Al/Fe]$<1.05$
(green symbols) and stars with [Al/Fe]$>1.05$ dex.
We see that (apart from a few outliers, most probably attributable to
uncertainties  both in the abundances and in the photometry) the segregation is
strikingly evident: the three structures traced by photometry along the RGB in
NGC~6752 nicely correspond to first generation stars in a tight sequence to the
left of the RGB and to second generation stars separated into two discrete
groups with moderate and very high nuclear processing, as indicated by their Al
abundances.

\section{A cluster analysis}

To better characterize the three different populations we found in NGC~6752, we
applied a cluster analysis to our data, like that
in \cite{gratton11} for $\omega$~Cen, using the $k-$means algorithm
\citep{Steinhaus,MacQueen} as implemented in the $R$\  statistical package (R
Development Core Team, 2011)\footnote{http://www.R-project.org}.
We considered five parameters for each star: [Na/Fe], [Mg/Fe], [Al/Fe], [Si/Fe],
and the residuals $\Delta(\delta_4$) of $\delta_4$\ values with respect to a
mean line drawn through the points for~NGC6752 in the ($y$, $\delta_4$) diagram.
The parameters were  weighted according to their internal errors. A complete set
of these  parameters is available for 71 stars: the cluster analysis was done on
these and we looked for a solution with three groups. The final division is
actually dominated by the Al abundances, which have the largest variation with
respect to internal errors. We called the three groups P, I, and E, because they
look similar to the subdivision in \cite{car09a}. In this sample, 15 stars
are classified P, 26 I, and 30 E. We will not discuss here these ratios, but
rather focus on the average abundance we obtain for the three different groups,
listed in Table~2.

The three groups order consistently according to all parameters. Perhaps the
most remarkable issue concerns the Mg and O abundances. If the I population
simply formed by a fraction $d$\ if the material of the E population diluted by
a fraction $(1-d)$\ of the material from the P population, the following
relation should hold:
\begin{equation}
d = { {[A({\rm Mg})_I-A({\rm Mg})_P]} \over {[A({\rm Mg})_E-A({\rm Mg})_P]} } = { {[A({\rm O})_I-A({\rm O})_P]} \over {[A({\rm O})_E-A({\rm O})_P]} }
\end{equation}
Instead, we get two very different values of $d$\ from O and Mg
($0.73\pm 0.13$ vs. $0.20\pm 0.18$). This fact is indeed confirmed by a
simple look at the relation between [O/Fe] and [Mg/Fe] (see Fig.~\ref{f:fig2}),
which is quite obviously not linear.

We conclude that the material from which the I population formed was not
produced by diluting material of the E population with material from the P
population. Mg is not depleted in the I population as much as expected from this
assumption, if the dilution is derived from O. The conclusion is that the
polluters for the E and I populations were different, the first one requiring
H-burning at higher temperature than the second.

\section{Discussion and conclusions}

In the present work we were able to provide a chemical tagging of three distinct
sequences of stars along the RGB in NGC~6752, using the largest set of
abundances of O, Na, Mg, Al, and Si for giants in this cluster.

The formation scenario and the nature of the polluters are different if
the abundances can be described by a dilution model (see e.g. \citealt{pra06})
rather than by a number of discrete populations, each characterized by a given
chemical composition. We examine our set of abundances from this point of view.

We provide two independent proofs that the distribution of giants in NGC~6752 is
multi-modal, with stars segregated into three distinct groups, according to both
Str\"omgren photometry and high-resolution spectroscopy. All these observations
are based on the different composition presented by RGB stars. The large
statistics from the photometry and the precise chemical tagging from our
abundance analysis allow to reliably conclude that the groups showing a
different chemistry are actually distinct.

As for the nature of the observed discreteness, it is easy to identify the 
first separation  with the gap in star formation occurring at very early times
between first and second generation stars. 

The second gap, observed among second generation stars, opens some questions. 
The definition of I and E groups of second generation stars,
introduced in \cite{car09a} from Na, O abundances (at [O/Na]$=-0.9$ dex), is not
entirely arbitrary, resting on the long tails observed in [O/Na] ratios in some
clusters like NGC~2808 or M~5, and returns fractions of I and E stars that
nicely correlates with physical global parameters of GCs, like the total mass.
In the case of NGC~6752, this neat separation, not immediately evident from the
Na-O anticorrelation, is quite clear when Al abundances are considered. 

These results tell us that in NGC~6752 the interplay between star formation and
dilution with pristine material acted so that two groups of stars with rather
homogeneous chemical composition stand out. Thus, any mechanism resulting in
discrete clumping must include an interruption either in the star formation or
in the inflow of diluting material.

In addition, comparing O and Mg abundances we found that the composition of 
the I population cannot be obtained by diluting the material from which the E
population was formed with those from which the P population originated.
This calls for different polluters for the I and E populations.

This can be obtained in various ways. In the case of AGB stars
we may think of two star formation episodes, separating the outcome of more
massive and less massive AGB stars. An analogous scenario for FRMS
is quite contrived, given their short lifetimes.
 
Another possibility is that both FRMS and AGB do contribute, at different
epochs, to modify the primordial composition of stars in a GC, as
suggested by Valcarce  \& Catelan (2011).
Were this the case, our observations require that the beginning of secondary
star formation did not take place in the low mass tail of first generation
stars: there must be a hiatus, albeit limited in time, between the two, else we
should observe only two groups, not three. 

Hence, we could envisage that the first contributors were FRMS that, owing to
their very large masses, started to pollute the intracluster environment in the
first few millions years after the burst of primordial star formation. After all
the remaining gas has been either transformed into stars or blown away by SNe,
no relevant amount of polluted matter was released with sufficiently
slow velocity to be retained in the cluster. Only after about  50-100 Myr, when
stars with an initial mass of about 7-5 $M_\odot$ entered  the AGB phase,
another injection of nuclearly processed matter occurred. Were
the surface density of the accumulated gas above a certain threshold, star
formation could be restarted forming stars with a chemistry of second
generation clearly distinct from the progeny of FRMS.

This is only a qualitative sketch that, however, might account with sake of
hypothesis for the existence of three separated populations of stars along the
RGB in NGC~6752, although our data do not allow to attribute each individual 
group to a given class of polluters. This analysis might be applied
to more metal-rich clusters. However, in these cases a careful
examination of the possible contamination from CN features (scarcely relevant in
the metal-poor cluster NGC~6752) must be included.

\acknowledgements
Partial funding comes from the PRIN INAF 2009 grant CRA 1.06.12.10 (PI: R.
Gratton).

\clearpage
\begin{deluxetable}{rrccrccrcccc}
\tabletypesize{\scriptsize}
\tablecaption{[Mg/Fe], [Al/Fe] and [Si/Fe] ratios for red giants in NGC~6752. 
Complete table available on line only.\label{abu}}
\tablehead{\colhead{star}&
\colhead{nr}&
\colhead{[Mg/Fe]}&
\colhead{$\sigma$}&
\colhead{nr}&
\colhead{[Al/Fe]}&
\colhead{$\sigma$}&
\colhead{nr}&
\colhead{[Si/Fe]}&
\colhead{$\sigma$}&
\colhead{flagAl}&
\colhead{flagPop}\\
}
\startdata
 2162  &1 & 0.452 &	&2 & 1.204 &0.106  & 8  & 0.494 &0.182 &  1  & -\\
 4602  &1 & 0.154 &	&2 & 1.184 &0.040  & 7  & 0.527 &0.189 &  1  & E \\
 4625  &2 & 0.423 &0.051&2 &-0.098 &0.008  & 8  & 0.401 &0.144 &  1  & P \\
 4787  &2 & 0.384 &0.093&2 & 1.150 &0.082  &11  & 0.491 &0.214 &  1  & E \\
\enddata
\tablenotetext{a}{nr is the number of lines.}
\tablenotetext{b}{$\sigma$ is the $rms$ scatter of the mean.}
\tablenotetext{c}{flagAl: 1 for detection and 0 for upper limits in Al}
\tablenotetext{d}{flagPop: P, I, or E as derived from the cluster analysis}
\end{deluxetable}

\begin{deluxetable}{cccccccc}
\tabletypesize{\scriptsize}
\tablecaption{Average values of parameters for stars in the three populations
of red giants in NGC~6752.}
\tablehead{\colhead{Pop.}&
\colhead{nr}&
\colhead{$\Delta(\delta_4$)} &
\colhead{[O/Fe]}&
\colhead{[Na/Fe]}&
\colhead{[Mg/Fe]}&
\colhead{[Al/Fe]}&
\colhead{[Si/Fe]}\\
}
\startdata
P & 15 &$-0.048\pm 0.006$&$+0.54\pm 0.04$&$+0.01\pm 0.03$&$+0.45\pm 0.02$&$+0.11\pm 0.04$&$+0.51\pm 0.01$\\
I & 26 &$-0.002\pm 0.003$&$+0.25\pm 0.04$&$+0.36\pm 0.03$&$+0.42\pm 0.01$&$+0.74\pm 0.04$&$+0.48\pm 0.01$\\
E & 30 &$+0.027\pm 0.003$&$+0.06\pm 0.04$&$+0.62\pm 0.02$&$+0.31\pm 0.01$&$+1.21\pm 0.02$&$+0.43\pm 0.01$\\
\enddata
\end{deluxetable}

\clearpage

\clearpage

\clearpage

\end{document}